\documentclass{amsart}

\usepackage{amsmath,amsthm}

\title{A Non-linear Schr\"odinger Type Formulation of FLRW Scalar Field Cosmology}

\author{Jennie D'Ambroise}
\address{Department of Mathematics and Statistics, University of
  Massachusetts, Amherst, MA 01003, USA}
\email{dambroise@math.umass.edu}

\author{Floyd L. Williams}
\address{Department of Mathematics and Statistics, University of
  Massachusetts, Amherst, MA 01003, USA}
\email{williams@math.umass.edu}


\newtheoremstyle{theorem}
{10pt} 
{10pt} 
{\sl} 
{\parindent} 
{\bf} 
{. } 
{ } 
{} 
\theoremstyle{theorem}
\newtheorem{theorem}{Theorem}

\newtheoremstyle{defi}
{10pt} 
{10pt} 
{\rm} 
{\parindent} 
{\bf} 
{. } 
{ } 
{} 
\theoremstyle{defi}

\begin{document}

\maketitle

\begin{abstract}
We show that the Friedmann-Lema\^itre-Robertson-Walker equations with scalar
field and perfect fluid matter source are equivalent to a suitable
non-linear Schr\"odinger type equation.  This provides for an alternate
method of obtaining exact solutions of the Einstein field equations for a
homogeneous, isotropic universe.
\end{abstract}

\section{Introduction}
\setcounter{equation}{0}
Recently, there have been interesting reformulations of Einstein field
equations for scalar field cosmologies (both for isotropic and anisotropic
models) in terms of generalized types of Ermakov-Milne-Pinney (EMP)
equations; see ~\cite{1}, ~\cite{3}, ~\cite{4}, ~\cite{8}, for example.  Such equations occur in a
variety of physical contexts and in particular have served to provide a
link between gravitational and non-gravitational systems ~\cite{5}.  We present
in this paper an alternate (non-EMP) formulation of homogeneous, isotropic
scalar field cosmology.  Namely, we provide a formulation in terms of a
non-linear Schr\"odinger type equation.  Some applications to exact field
solutions are presented, including some string-inspired cosmological
solutions.  

One can set up a direct correspondence between EMP solutions and
Schr\"odinger solutions which suggests, for example, that possible
connections between this work and that in~\cite{5} can be pursued.  The
Schr\"odinger equation here also seems to be of some independent interest.

\section {Einstein Equations}
\setcounter{equation}{0}
The Einstein equations for a Friedmann-Lema\^itre-Robertson-Walker (FLRW)
homogeneous, isotropic universe with scalar field $\phi$, potential $V$,
and perfect fluid matter source assume the familiar form (for a vanishing
cosmological constant)

\begin{equation}
H^2+\frac{k}{a^2} \stackrel{(i)}{=} \frac{8\pi G}{3}\rho_T\,, \qquad
\frac{\ddot{a}}{a} \stackrel{(ii)}{=}\frac{-4\pi G}{3}\left(\rho_T+3p_T\right),
\end{equation}
where the total energy density $\rho_T$ and pressure $p_T$ with matter
contributions $\rho_m$, $p_m$ are given by 

\begin{equation}
\rho_T=\rho_\phi+\rho_m\,, \qquad p_T=p_\phi+p_m
\end{equation}
for
\begin{align}
\rho_\phi & = \frac{\dot{\phi}^2}{2}+V\circ\phi\,,& p_\phi &=
\frac{\dot{\phi}^2}{2}-V\circ \phi\,, \\
\rho_m    & = \frac{D}{a^n}\,,& p_m&=\frac{(n-3)D}{3a^n}
\end{align}
with $D$, $n$ = constants $\geq \,0$, $n\neq 0$,
$H\stackrel{def.}{=}\frac{\dot{a}}{a}=$ the Hubble parameter for the scale
factor $a=a(t)$, and $k=0$, 1, or $-1$ = the curvature parameter.  $G$ is
Newton's constant and units are selected so that the speed of light is
unity; cf.~\cite{7}.  Equations (i), (ii) imply the fluid conservation equation
\begin{equation}
\dot{\rho_T}+3H(\rho_T+p_T)=0,
\end{equation}
which by definitions (2.3), (2.4) reduces to the Klein-Gordon equation
of motion
\begin{equation}
\ddot{\phi}+3H\dot{\phi} +V'\circ \phi =0
\end{equation}
of the scalar field $\phi$.  Note that for
$\gamma_m\stackrel{def.}{=}\frac{n}{3}$ one has the equation of state
$p_m=(\gamma_m-1)\rho_m$ (by (2.4)).  For $\gamma_m=1$ (i.e. n=3), for
example, $p_m=0$, which is the case of a \underline{dust} universe.

There is also an equation of state $p_\phi=(\gamma_\phi-1)\rho_\phi$ which
follows by setting $\gamma_\phi=2\dot{\phi}^2\left[
  \dot{\phi}^2+2(V\circ\phi)\right]^{-1}$.  $\gamma_\phi$ however, unlike
$\gamma_m$, is a non-constant function of time $t$.

In this paper we set up a correspondence $(a,\phi, V)\longleftrightarrow u$
between solutions $(a,\phi, V)$ of the field equations (i), (ii) in (2.1)
and solutions $u$ of the following time-independent Schr\"odinger type
equation
\begin{equation}
u''(x)+\left[E-P(x)\right]u(x) = -\frac{nk}{2}u(x)^{\frac{4-n}{n}}
\end{equation}
with constant energy $E$ and potential $P(x)$.  For a flat universe ($k=0$),
or for the special values n=2 or 4 one notes that equation (2.7) is
actually a \underline{linear} Schr\"odinger equation.  The correspondence
provides an alternate method, for example, of solving the equations in
(2.1).  In some cases it is convenient to specify the scale factor $a(t)$ a
priori and then find the scalar field $\phi$ and potential $V$ such that
$(a, \phi , V)$ is a solution in (2.1).  This can be done, in particular,
in our approach as $a(t)$ is sufficient to determine $u(x)$ in (2.7), which
by the correspondence determines the desired $\phi$ and $V$.

\section{Description of The Correspondence $(a, \phi ,V) \longleftrightarrow u$}
\setcounter{equation}{0}
With the above notation in place, and where we set $K^2 \stackrel{def.}{=}
8\pi G$ for convenience, we can state the main theorem:
\begin{theorem}
Let $u(x)$ be a solution of equation (2.7), given $E$, $P(x)$.  Then a
solution $(a, \phi ,V)$ of the Einstein equations (i), (ii) in (2.1) can be
constructed as follows, where $D$ in (2.4) (which specifies $\rho_T$, $p_T$)
is chosen to be $\frac{-12E}{n^2K^2}$: First choose functions $\sigma(t)$,
$\psi(x)$ such that
\begin{equation}
\dot{\sigma}(t) =u(\sigma(t))\,, \quad \psi '(x)^2 =\frac{4}{nK^2}P(x). 
\end{equation}
Then one can take
\begin{eqnarray}
a(t) &=& u(\sigma(t))^{-\frac{2}{n}}, \qquad \phi(t) = \psi(\sigma(t)), \\
V &=& \left[ \frac{12}{K^2n^2}(u')^2-\frac{2u^2P}{K^2n}+\frac{12u^2E}{K^2n^2}+\frac{3ku^{\frac{4}{n}}}{K^2} \right]\circ \psi ^{-1}.
\end{eqnarray}
Here, in fact, $(a, \phi , V)$ will also satisfy the equations
\begin{eqnarray}
\dot{\phi}(t)^2 &=& \frac{-2}{K^2}\left[ \dot{H}(t)-\frac{k}{a(t)^2} \right]-\frac{nD}{3a(t)^n}\\
V(\phi(t)) &=& \frac{3}{K^2}\left[ H(t)^2+\frac{\dot{H}(t)}{3}+\frac{2k}{3a(t)^2} \right]+\frac{(n-6)D}{6a(t)^n}\mbox{;}
\end{eqnarray}
again $D\stackrel{def.}{=}\frac{-12E}{n^2K^2}$.  Conversely, let $(a,
  \phi , V)$ be a solution of equations (i), (ii) in (2.1), with some $D$
  given in (2.4) that specifies $\rho_T$, $p_T$ in (2.2).  Similar to
  (3.1), choose some solution $\sigma(t)$ of the equation
\begin{equation}
\dot{\sigma}(t)=a(t)^{-\frac{n}{2}}.
\end{equation}
Then equation (2.7) is satisfied for
\begin{eqnarray}
E &\stackrel{def.}{=}& -\frac{K^2n^2}{12}D\,, \\
P(x) &\stackrel{def.}{=}& \frac{nK^2}{4}a(\sigma^{-1}(x))^n\left[\dot{\phi}(\sigma^{-1}(x))\right]^2\,,\\
u(x) &\stackrel{def.}{=}& a(\sigma^{-1}(x))^{-\frac{n}{2}}.
\end{eqnarray}
\end{theorem}
\noindent Theorem 1 therefore provides for a concrete correspondence $(a, \phi , V)
\leftrightarrow u$ between solutions $(a, \phi , V)$ of the gravitational
field equations (i), (ii) in (2.1) and solutions $u$ of the non-linear
Schr\"odinger type equation (2.7).  The solutions $(a, \phi , V)$ also
correspond to solutions $Y$ of the generalized Ermakov-Milne-Pinney
equation
\begin{equation}
Y''+QY = \frac{\lambda}{Y^3}+\frac{nk}{2Y^{\frac{(n+4)}{4}}}\,,
\end{equation}
as discussed in ~\cite{8}; also see ~\cite{1}, ~\cite{4}.  In turn, one can set up a correspondence
$Y \leftrightarrow u$ between solutions $Y$ of (3.10) and solutions $u$ of
(2.7), and thus obtain the correspondence $(a, \phi , V)\leftrightarrow u$
-- and the proof of Theorem 1.  However, we prefer to give a
\underline{direct} proof of Theorem 1 -- one that does not rely on the
results in ~\cite{8}, nor on the correspondence $Y \leftrightarrow u$ -- although
the latter route served as the motivation for the formulation of equation
(2.7), and consequently of Theorem 1.  It is implicitly assumed in
definition (3.3) that the inverse function $\psi ^{-1}$ of $\psi$ exists,
which would not be the case if $P(x)=0$.  Therefore we first assume that
$P(x)$ is not the zero function.  The case $P(x)=0$ will be discussed
later.

For the proof of Theorem 1 we first establish the salient formulas (3.4),
(3.5), given $E$, $P(x)$, $u(x)$ in (2.7).  By (3.2), $u\circ \sigma
\stackrel{def.}{=} a^{-\frac{n}{2}}$ which differentiated in conjunction with (3.1) gives
$(u' \circ \sigma)(u \circ \sigma) = -\frac{n}{2} a^{-\frac{n}{2}-1}\dot{a}\stackrel{def.}{=} -\frac{n}{2}(u \circ
\sigma)H$.  That is, $u' \circ \sigma = -\frac{n}{2}H$ which we also
differentiate to obtain $(u''\circ \sigma)(u \circ\sigma)=-\frac{n}{2}\dot{H}$ (again by (3.1)); i.e.,
\begin{equation}
u \circ \sigma = a ^{-\frac{n}{2}}\,, \quad u'\circ \sigma =
-\frac{n}{2}H\,, \quad (u''\circ \sigma)(u\circ \sigma)=-\frac{n}{2}\dot{H}.
\end{equation}
From equation (2.7), $Pu^2=u''u+Eu^2+\frac{nk}{2}u^{\frac{4}{n}}$ so that
by (3.11)
\begin{equation}
(Pu^2)\circ \sigma =-\frac{n}{2}\dot{H} + Ea^{-n} +\frac{nk}{2}a^{-2}.
\end{equation}
Using (3.11), (3.12) we now obtain by definitions (3.2), (3.3),
\begin{eqnarray*}
V\circ \phi &=&
\frac{12}{K^2n^2}\left(-\frac{n}{2}H\right)^2-\frac{2}{K^2n}\left(-\frac{n}{2}\dot{H}+Ea^{-n}+\frac{nk}{2}a^{-2}\right)+\frac{12E}{K^2n^2}a^{-n}+\frac{3k}{K^2}a^{-2}\\
&=&
\frac{3H^2}{K^2}+\frac{\dot{H}}{K^2}+\frac{D}{a^n}\left(\frac{n}{6}-1\right)+\frac{2k}{K^2}a^{-2} \,\,\,\,
(\mbox{since} D\stackrel{def.}{=} -\frac{12E}{n^2K^2}),
\end{eqnarray*}
 which is equation (3.5) as desired.  Again by (3.1), (3.2), $\dot{\phi}=(\psi ' \circ \sigma)(u\circ
\sigma) \Rightarrow \dot{\phi}^2\stackrel{def.}{=} \frac{4}{nK^2}(P\circ \sigma)(u \circ \sigma)^2 =$ (by (3.12))
$\frac{4}{nK^2}\left(-\frac{n}{2}\dot{H}+Ea^{-n}+\frac{nk}{2}a^{-2}
\right)=-\frac{2}{K^2}\dot{H}-\frac{nD}{3}a^{-n}+\frac{2k}{K^2}a^{-2}$, which is equation (3.4).

With equations (3.4), (3.5) now established, we can compute $\rho_T$, $p_T$
in (2.2), using (2.3):
\begin{eqnarray*}
\rho_\phi &=&
-(\dot{H}-ka^{-2})K^{-2}-\frac{nD}{6}a^{-n}+3(H^2+\frac{\dot{H}}{3}+\frac{2k}{3}a^{-2})K^{-2}\\
& &\,\,\,+\frac{(n-6)}{6}Da^{-n}\,\,\, (\mbox{by} (2.3), (3.4), (3.5))\\
&=& (3ka^{-2}+3H^2)K^{-2}-Da^{-n} \\
& \stackrel{def.}{=}& (3ka^{-2}+3H^2)K^{-2}-\rho_m \,\,\,\,\, (\mbox{by}
(2.4))\\ 
\Rightarrow \rho_T &\stackrel{def.}{=}& \rho_\phi+\rho_m \\
& \stackrel{(i)'}{=}& 3(ka^{-2}+H^2)K^{-2},
\end{eqnarray*}
 which is exactly equation (i) in (2.1).  Similarly, by (2.3), (3.4), (3.5),
\begin{eqnarray*}
p_\phi &=&
-(\dot{H}-ka^{-2})K^{-2}-\frac{n}{6}Da^{-n}-3(H^2+\frac{\dot{H}}{3}+\frac{2k}{3}a^{-2})K^{-2}-\frac{(n-6)}{6}Da^{-n} \\
&=& -2\dot{H}K^{-2}-ka^{-2}K^{-2}-3H^2K^{-2}+\frac{(3-n)}{3}Da^{-n} \\
&{\stackrel{def.}{=}}& -(2\dot{H}+ka^{-2} +3H^2)K^{-2}-p_m \,\,\,(\mbox{by} (2.4)) \\
\Rightarrow p_T &\stackrel{def.}{=}&
p_\phi+p_m=-(2\dot{H}+ka^{-2}+3H^2)K^{-2},
\end{eqnarray*}
 which with $(i)'$ gives
$\rho_T+3p_T=3(ka^{-2}+H^2)K^{-2}-3(2\dot{H}+ka^{-2}+3H^2)K^{-2}=-6(H^2+\dot{H})K^{-2}=-6\frac{\ddot{a}}{a}K^{-2}$,
which is exactly equation $(ii)$ in (2.1).  The arrow $u \longrightarrow
(a, \phi , V)$ has therefore been established in one direction.

For the other direction $(a, \phi , V) \longrightarrow u $, with $(a,
\phi,V)$ a solution of the equations in (2.1), we must show that $u(x)$
defined in (3.9) solves equation (2.7), for the data $E$, $P(x)$ defined in
(3.7), (3.8), with $\sigma(t)$ defined by (3.6).  For convenience let $g(x)
= \sigma ^{-1} (x)$ denote the inverse function of $\sigma(t)$:
$\sigma(g(x))=x \Rightarrow \dot{\sigma}(g(x))g'(x)=1$.  That is,
\begin{equation}
u(x)g'(x) =1
\end{equation}
since by (3.6), (3.9) this product is $a(g(x))^{-\frac{n}{2}}g'(x)=
\dot{\sigma}(g(x))g'(x)$.  Then
\begin{eqnarray*}
u'(x)&=&-\frac{n}{2}a(g(x))^{-\frac{n}{2}-1}\dot{a}(g(x))g'(x) \\
&\stackrel{def.}{=}& -\frac{n}{2}u(x)H(g(x))g'(x)\\
& =& -\frac{n}{2}H(g(x))\\
\Rightarrow u''(x) &=& -\frac{n}{2}\dot{H}(g(x))g'(x)\\
&=& -\frac{n}{2}\left[\frac{\ddot{a}(g(x))}{a(g(x))}-H(g(x))^2\right]g'(x),
\end{eqnarray*}
where by (2.1) the bracket here is
\begin{eqnarray*}
&& -\frac{K^2}{6}\left[ \rho_T(g(x))+3p_T(g(x))
\right]+\frac{k}{a(g(x))^2}-\frac{K^2}{3}\rho_T(g(x))\\
&=& -\frac{K^2}{2}\left[  \rho_T(g(x))+p_T(g(x))
\right]+ku(x)^{\frac{4}{n}}
\,\,\, (\mbox{i.e. }\,\,u(x)^{\frac{-2}{n}}\stackrel{def.}{=}a(g(x)))\\
&=& -\frac{K^2}{2}\left[\dot{\phi}(g(x))^2+\frac{nD}{3a(g(x))^n}\right]+ku(x)^{\frac{4}{n}}\,\,\,\, (\mbox{by}
(2.2), (2.3), (2.4)) \\
&=& -\frac{K^2}{2}\left[\frac{4P(x)}{nK^2}u(x)^2+\frac{nD}{3}u(x)^2
\right]+ku(x)^{\frac{4}{n}},\,\,\mbox{by} (3.8), (3.9).
\end{eqnarray*}
That is, we see that
$u''(x)=-\frac{n}{2}g'(x)\left\{-\frac{K^2}{2}\left[\frac{4P(x)}{nK^2}u(x)+\frac{nD}{3}u(x)\right]u(x)+ku(x)^{\frac{4}{n}-1}u(x)\right\}=P(x)u(x)+\frac{n^2K^2}{12}Du(x)-\frac{n}{2}
ku(x)^{\frac{4}{n}-1}$ (by (3.13)), which by (3.7) is exactly equation
(2.7).  The proof of Theorem 1 is therefore complete, where we have assumed
the existence of $\psi^{-1}$ in (3.3).

\underline{Remarks.}
\begin{enumerate}
\item The case $P(x)=0$:\\
 In equation (2.7) we generally take a non-zero
  potential $P(x)$.  If $P(x)=0$ then $\psi(x)$ is a constant function by
  (3.1) and therefore its inverse function $\psi^{-1}(x)$ in definition
  (3.3) does \underline{not} exist, which means that the expression for $V$
  there has no meaning.  However, if $P(x)=0$ we can instead define $V(x)$
  to be a constant function and we also define $\phi(t)$ to be a constant
  function, because of (3.2).  First note that equation (3.4) still holds
  (with the left hand side there being 0 of course) since by (3.12),
  $0=-\frac{n}{2}\dot{H}-\frac{n^2K^2}{12}Da^{-n}+\frac{nk}{2}a^{-2}$ (again
  as $D\stackrel{def.}{=}-\frac{12E}{n^2K^2}$), which multiplied by
  $\frac{4}{nK^2}$ gives
\begin{equation}
0=-\frac{2}{K^2}\dot{H}-\frac{nD}{3}a^{-n}+\frac{2k}{K^2}a^{-2},
\end{equation}
as claimed.  Next note that the right hand side of (3.5) indeed is a
constant function of $t$ (for ($P(x)=0$).  Namely, differentiate (3.14) to
obtain
\begin{equation}
\ddot{H}=\left(-\frac{2k}{a^2}+\frac{K^2n^2D}{6a^n}\right)H
\end{equation}
and then compute that
\begin{eqnarray*}
& & \frac{d}{dt}\left\{\frac{3}{K^2}\left[
    H^2+\frac{\dot{H}}{3}+\frac{2k}{3a^2} \right]+\frac{(n-6)}{6a^n}D
\right\}\\
&=& \frac{3}{K^2}\left[
    2H\dot{H}+\frac{\ddot{H}}{3}-\frac{4kH}{3a^2}\right]-\frac{(n-6)nDH}{6a^n}\\
&=& \frac{3}{K^2}\left[
    2H\dot{H}+\left(\frac{-2k}{3a^2}+\frac{K^2n^2D}{18a^n}\right)H-\frac{4kH}{3a^2}\right]
    -\frac{(n-6)nDH}{6a^n} \,\,\,\,(\mbox{by}\,\,(3.15))\\
&=& 3\left[\frac{2\dot{H}}{K^2}-\frac{2k}{K^2a^2}
    \right]H+\frac{n^2D}{6a^n}H-\frac{n^2DH}{6a^n}+\frac{nDH}{a^n} \\
&=&  3\left(\frac{-nD}{3} \right)a^{-n}H+\frac{nDH}{a^n}\,\,\,\,(\mbox{by}\,\,(3.14)) \\
&=& 0
\end{eqnarray*}
$\Rightarrow$ (as claimed) that for some constant $V_0$ one has
\begin{equation}
\frac{3}{K^2}\left[H(t)^2+\frac{\dot{H}(t)}{3}+\frac{2k}{3a(t)^2}\right]+\frac{(n-6)D}{6a(t)^n}=V_0.
\end{equation}
In summary, in case $P(x)=0$ in (2.7) we define $\phi(t)=$ any constant and
$V(x)=V_0$ in (3.16) (as definition (3.3) no longer has a meaning).  Then
equations (3.14), (3.16) (the proper versions of equations (3.4), (3.5))
hold.\\

It follows that, again with
$a(t)\stackrel{def.}{=}u(\sigma(t))^{-\frac{2}{n}}$, where
$\dot{\sigma}(t)=u(\sigma(t))$, one does arrive at a solution $(a , \phi ,
V)$ of (i), (ii), in (2.1).  Namely, in (2.2), (2.3), (2.4), $\rho_\phi(t) = V_0$,
 $p_\phi(t)= -V_0$,
\begin{eqnarray*}
\Rightarrow \rho_T &=& V_0 +Da^{-n}\\
& =& \frac{3}{K^2}\left[ H^2+\frac{\dot{H}}{3}+\frac{2k}{3a^2}
\right]+\frac{nD}{6a^n}\,\,\, (\mbox{by}\,\, (3.16))\\
&=& \frac{3}{K^2}\left[H^2+\frac{K^2}{6}\left(-\frac{nDa^{-n}}{3}+\frac{2k}{K^2}a^{-2}
  \right)+\frac{2k}{3a^2} \right] +\frac{nD}{6a^n}\,\,\,\, (\mbox{by}\,\, (3.14)) \\
&=& \frac{3}{K^2}\left[ H^2+\frac{k}{a^2}\right],
\end{eqnarray*}
 which is (i), and
\begin{eqnarray*}
p_T&=&-V_0+\frac{(n-3)D}{3}a^{-n} \\
\Rightarrow \rho_T+3p_T &=& -2V_0+(n-2)Da^{-n}\\
&=& -\frac{6}{K^2}\left[
  H^2+\frac{\dot{H}}{3}+\frac{2k}{3a^2}\right]-\frac{(n-6)D}{3a^n}+(n-2)Da^{-n}\,\,\, (\mbox{by} (3.16))\\
&=& -\frac{6}{K^2}\left[ H^2+\frac{\dot{H}}{3}
\right]-\frac{4k}{K^2a^2}+\frac{2Dn}{3a^n}\\
&=&-\frac{6}{K^2}\left[H^2+\frac{\dot{H}}{3}\right]+2\left(
  -\frac{2}{K^2}\dot{H}\right)\,\,\, (\mbox{by} (3.14))\\
&=& -\frac{6}{K^2}\left[H^2+\dot{H}\right]\\
&=& -\frac{6}{K^2}\frac{\ddot{a}}{a},
\end{eqnarray*} which is (ii). 

\item Equations (3.4), (3.5) imply (i), (ii):\\
The argument following the establishment of equations (3.4), (3.5) actually
shows (independently of $u(x)$) that if $a(t)$ is given a priori, and if
$\phi(t)$, $V(x)$ are functions satisfying equations (3.4), (3.5), then
automatically $(a,\phi , V)$ solves equations (i) (ii) in (2.1).
\end{enumerate}

\section{Some Examples}
\setcounter{equation}{0}
As a simple illustration of the application of Theorem 1, choose
$u(x)=\left(1-A^4\omega^2x^2 \right)/A^2$ for $A,\omega >0$, which solves
equation (2.7) for $E=0$,
$P(x)=2A^2\left(1-A^2\omega^2\right)\left(1-A^4\omega^2x^2\right)^{-1}=K^2B^2A^4\left(1-A^4\omega^2x^2\right)^{-1}$,
$B^2=2(1-A^2\omega^2)/K^2A^2$, $n=4$, $k=1$.  In (3.1) we can take
$\sigma(t)=\tanh(\omega t)/A^2\omega$, $\psi(x)=\psi _0 \pm
\frac{B}{\omega}\arcsin(A^2\omega x)=\psi _0 \pm \frac{B}{\omega}\arctan
(\frac{A^2\omega x}{\sqrt{1-A^4\omega^2x^2}})$ for $A^2\omega |x|<1$ and
obtain from (3.2), (3.3), $a(t)=A\cosh(\omega t)$, $\phi(t)=\psi _0 \pm
\frac{B}{\omega}\arcsin(\tanh(\omega t)) =\psi _o
\pm\frac{B}{\omega}\arctan(\sinh (\omega t))=\psi _0 '\pm
\frac{2B}{\omega}\arctan(e^{\omega t})$,
$V(x)=\frac{3\omega^2}{K^2}+B^2\cos ^2 \left(\frac{\omega}{B}(x-\psi
  _0)\right)$, for constants $\psi _0$, $\psi _0 '$, which is the
Ellis-Madsen solution in section 4.3 of ~\cite{2}.  We note that the 2 in the
expression $\sin\left(2\frac{\omega}{B}(\phi - \phi_0) \right)$ in equation
(42) of ~\cite{2} should not appear there.  One can obtain in fact all of the
solutions in ~\cite{2} for a suitable choice of $u(x)$ and $n$.

\"Ozer and Taha have considered in ~\cite{6} two string-motivated solutions $(a_j,
\phi _j , V_j)$, $j=1,2$, for $k=1$, $D_j=0$, where the potentials $V_j$
were specified a priori.  One can also take the point of view of specifying
the scale factors $a_1(t)=(a_0^2+t^2)^{\frac{1}{2}}$,
$a_2(t)=a_0+t^2/2a_0$, for $a_0\neq 0$, and then applying Remark 2 to find
$(\phi _j , V_j)$.  For $a_2(t)$, for example, $H_2(t)=2t/(2a_o^2+t^2)$,
$\dot{H_2}(t)=2(2a_0^2-t^2)/(2a_0^2+t^2)^2$, and one obtains by (3.2)
$\dot{\phi_2}(t)^2=4t^2/K^2(2a_0^2+t^2)^2 \Rightarrow
\phi_2(t)=\phi_0+\frac{1}{K}\log\left(1+\frac{t^2}{2a_0^2} \right)$, where
we choose the positive square root.  Also,
$H_2(t)^2+\dot{H_2}(t)^2/3+2/3a_2(t)^2=(10t^2+12a_0^2)/3(2a_0^2+t^2)^2
\Rightarrow V_2(\phi_2(t))=(10t^2+12a_0^2)/K^2(2a_0^2+t^2)^2$, by (3.5).
For $x>\phi_0$, we can take
$\phi_2^{-1}(x)=\sqrt{2}|a_0|\left[e^{K(x-\phi_0)}-1 \right]^{\frac{1}{2}}$
and compute that $V_2(x)=V_2(\phi_2(\phi_2^{-1}(x)))=\left[5e^{-K(x-\phi_0)}-2e^{-2K(x-\phi_0)}
\right]/K^2a_0^2$.  Similarly, for the above scale factor $a_1(t)$ one can
obtain via (3.4), (3.5) $\phi _1(t)=\phi
_0'+\frac{1}{K}\log\left(1+\frac{t^2}{a_0^2}\right)$,
$V(x)=\left[4e^{-K(x-\phi_0 ')}-e^{-2K(x-\phi_0 ')} \right]/K^2a_0^2$, say
for $x>\phi _0 '$.  These solutions $(a_j, \phi _j, V_j)$ can also be
obtained by taking $u_j(x)=a_j(\sigma^{-1}(x))^{-\frac{n_j}{2}}$ (which is
motivated by (3.2)) for the convenient choices $n_1=4$, $n_2=2$, and using
the corresponding solutions $\sigma_1(t) =a_0^{-1}\arctan(a_0^{-1}t)$,
$\psi_1(x)=\psi_0-\frac{2}{K}\log(\cos(a_0x))$,
$\sigma_2(t)=\sqrt{2}\arctan((\sqrt{2}a_0)^{-1}t)$,
$\psi_2(x)=\psi_0-\frac{2}{K}\log(\cos(\frac{x}{\sqrt{2}}))$ of the
  equations in (3.1) for $P_1(x)=4a_0^2\tan ^2(a_0x)$, $P_2(x)=\tan
  ^2(\frac{x}{\sqrt{2}})$; $E_1=E_2=0$.  One can go beyond the assumption
  $D_1=D_2=0$ (in ~\cite{6}) and obtain solutions $\phi_j$, $V_j\circ \phi_j$ via
  (3.4), (3.5).  For example, for $D_1 \neq 0$ one can check, using
  Mathematica for example, that
\begin{equation}
\phi_1(t) = \phi_0 ' + \frac{2}{\sqrt{3}K}\left\{
  \frac{-\sqrt{3a_0^2+DK^2}}{a_0}\mbox{arctanh}\left(\frac{\sqrt{3a_0^2+DK^2}t}{a_0\sqrt{-DK^2+3t^2}}\right)+\sqrt{3}\log\left( 3t +\sqrt{-3DK^2+9t^2}\right) \right\}
\end{equation}
and that
\begin{equation}
V_1(\phi_1(t)) = \frac{3(4t^2+3a_0^2)-DK^2}{3K^2(a_0^2+t^2)^2}.  
\end{equation}

As a fourth example, and final one, for a parameter $\lambda \in \mathbb{R}
\setminus \{0\}$, let $u(x)=-\frac{3}{4}\sqrt{3} \tanh
\left(\sqrt{\frac{27}{8}}\lambda x \right)$.  Then $u(x)$ solves equation
(2.7) for $E=A+\frac{27}{4}\lambda ^2$, $A>0$, $P(x)=A$, $n=1$,
$k=-8\lambda ^2$.  By an application of Theorem 1, one obtains the
solutions $(a, \phi_{\pm} , V)$ given by
\begin{eqnarray}
a(t) &=& \frac{16}{27}\left[ 1+ e^{\frac{27\lambda}{4\sqrt{2}}(t-c)}
\right], \notag \\
\phi_{\pm}(t) &=& \pm \frac{4}{3}\sqrt{\frac{2}{3}}\frac{\sqrt{A}}{\lambda
  K}\mbox{arcsinh}\left[ e^{-\frac{27\lambda}{8\sqrt{2}}(t-c)} \right] ,\\
V(x) &=& \frac{3^7\lambda ^2}{2^5K^2}+\frac{135A}{8K^2}\tanh ^2
\left[\frac{3}{4}\sqrt{\frac{3}{2}}\frac{\lambda K}{\sqrt{A}}(x-\phi_0) \right], \notag
\end{eqnarray}
for integration constants $c$, $\phi_0$.

\end{document}